# Quench dynamics of Wannier-Stark states in an active synthetic photonic lattice

*Alexander Dikopoltsev*[1,†], *Ina Heckelmann*[1,†], *Mathieu Bertrand*[1,†], *Mattias Beck*[1], *Giacomo Scalari*[1],

*Oded Zilberberg*[2], *and Jérôme Faist*[1]

[1]Physics Department, ETH Zurich, Switzerland

[2]Physics Department, Konstanz University, Germany

[†]These authors contributed equally

**Abstract**: Photonic emulators have enabled the study of a plethora of solid-state phenomena and contributed to the invention of optical quantum-inspired devices. Current photonic emulators are limited to bosonic behavior with local interactions, but the use of active synthetic lattices bears the possibility to overcome this limitation. In this work, we suggest the modulated ring fast-gain laser as a platform for emulation of quench dynamics in a synthetic lattice that follow equal density filling of their reciprocal space. As a demonstration of the strength of this emulation platform, we apply quenching onto a dispersed Wannier-Stark ladder and directly observe oscillations, enabled by the fast-gain, and their coherent stabilization to a single Wannier stark state. The coherent dynamics directly result from our laser's liquid state of light, a property that stems from fast-gain, explained by the fast decay of fluctuations that occur on the shortest timescale of the system. Moreover, we find that sufficient biasing of the lattice, induced by detuning the modulation from the cavity resonance, the process supports oscillatory dynamics in the synthetic space. This platform enriches our understanding of collective dynamics in the nonperturbative regime, but also improves our capabilities to control and generate coherent multi-frequency sources.

Analog computing is aimed at replicating complex phenomena without direct interaction with the original system. At its heart, a successful emulator requires tunable control operations as well as control over the initial boundary conditions of the simulation. Photonic emulators have emerged as a powerful platform for probing diverse solid-state phenomena, revealing complex quantum and classical dynamics. These emulators leverage on the properties of light, harnessing their extended coherence time due to typically weak interactions with the environment. This has led to the investigation of a plethora of phenomena such as Anderson localization[1], topological insulators[2], anyonic statistics[3], fractional Bloch oscillations[4], Thouless pumping[5–7] or moiré lattices[8] in a controlled and adjustable experimental setting. While part of the discoveries in these emulators were providing intuition about known solid state dynamics, the other portion could extend beyond the known physical models, for example by introducing non-Hermitian dynamics to the field[9], or by exploring models with dimensions larger than three[10]. Furthermore, photonic emulators not only provide the means for exploring fundamental aspects of physics, but also hold the potential to engender novel technologies and applications, e.g., in quantum information processing[11] or laser science[12–16].

The vast majority of photonic emulators are typically realized with spatially varying structures in real space. This poses constraints on dimensionality of the tested configurations, the range of available coupling schemes, and the type of nonlinearities for light in the specific system. To exceed the boundaries of real space, it is possible to construct dynamics in artificial spaces using other properties of light, such as time delay[17] or wavelength[10,18], effectively generating a new synthetic dimension. These synthetic dimensions allow us to overcome the fundamental limits that exist in realizations of lattices in physical space[19]. Systems that support dynamics in these artificial lattices provide a versatile platform for previously unattainable forms of settings and interactions. For example, in this manner, it is possible to increase the dimensionality of quantum systems[20] such as photonic crystals[21] or topological matter[10,22], or even control the non-Hermitian properties demonstrating PT-symmetric ballistic to diffusive transitions[23] or funneling of light[24].

The coherence of light stems from its limited inherent interaction with external fields. This leads to a hurdle in exploring interactions with gauge fields in photonic emulators. Luckily, the gauge fields can be artificially created through synthetic space design[19]. In the case of charge-like behavior, it is possible to introduce artificial electric and magnetic fields in the synthetic space for photons[18,25]. For example, coupled multimode resonators under collective resonant modulation lead to 2D synthetic lattices where interleaving of the modulation phase induces time-reversal symmetry breaking and appearance of topological edge states[26]. Similar methods were used with cold atoms to create magnetic fields in synthetic lattices of neutral particles[10,27,28]. Moreover, linear electrical gauge fields that induce Bloch oscillations dynamics were also reported in synthetic lattices[18,29,30]. The experiment necessitated the employment of an artificial field, generated by establishing a potential gradient within the lattice[31–33].

Amidst the vast research dedicated to light-coupling gauge fields, the influence of nonlinear interactions in synthetic dimensions has not been the topic of much investigation. Of the few pioneering works in this direction, a recent example demonstrated solitons that are exposed to a linear gauge field in synthetic space but preserve their oscillatory motion due to long-range interactions, caused by a local nonlinearity in real space[34]. This behavior significantly differs from short-range nonlinearities, typically caused by the Kerr effect in the same space of the wave dynamics, where Bloch oscillations experience self-focusing or self-defocusing[33], thus distorting the oscillation process. In another example, recently, a fast-gain laser showed the preservation of quantum-walk dynamics in synthetic space, leading to a ballistic expansion and stabilization of the laser spectrum, facilitated by the long-ranged nonlinearities of these types of devices[35]. Interestingly, the nonlinearity in these systems overcame a long existing limitation of mode-locked lasers, where dissipation induces decay and stabilization to the mode with the lowest bandwidth, i.e. the gaussian pulse[36], which arises from their bosonic nature[37–39]. These are key examples where comprehension and utilization of nonlinear phenomena in photonic lattices is critical not only for advancing our understanding of the naturally interacting electrons in lattices[40,41], but also for optical technologies[15,16].

Here, we propose and demonstrate that the modulated ring-cavity fast-gain laser acts as a photonic emulator of a synthetic lattice, with fast gain saturation forcing equal density in its reciprocal space. We let the laser pick an initial state from which quench dynamics in the system can be explored, exhibiting a rich competition between non-Hermitian dynamics along the synthetic dimension; in this context, the fast-gain nonlinearity acts against condensation, akin to collective fermion or hardcore boson statistics. Specifically, we use detuned modulation to induce coupling in the synthetic space alongside an artificial electrical field. This field and the dispersion in our system, also natural to biased electronic lattices, produce a dispersed Wannier-Stark ladder of modes around the initial state. The dispersion acts by corrupting the expected Bloch oscillations; however, the periodic dynamics still persist thanks to the equal filling of the reciprocal space, until the system stabilizes on one Wannier-Stark state. We show that this type of process occurs when the artificial electric field is strong enough to support oscillations that are not bound to the potential trap. Finally, we attribute the coherence of the oscillations and decay to the liquid-like state of the fast-gain system. This new photonic emulation method broadens our understanding of collective electronic dynamics in crystals.

**System and the synthetic space model**

The experimental platform of the emulator is based on the synthetic modal space of a ring cavity quantum cascade laser that supports multiple longitudinal modes[35], see Fig. 1A. Our laser device has insignificant backscattering and fast gain saturation[42,43], which together lead to single mode lasing. This sets the initial conditions for our emulator. At time $t_0$, we turn on a modulation through RF injection with amplitude $J_m$. When the modulation frequency matches the level spacing of the cavity's free spectral range $\Omega = 2.51 \cdot 10^9 \, rad/sec$, it induces coupling in a synthetic space spanned by the ring's modes[10,25,44,45], see Fig. 1B. Together, the dynamics in the copropagating frame, $z$, relative to the initial lasing mode is given by

$$i\dot{E} = ig\left(1 - \frac{I}{I_s}\right)E - i\alpha E + i\frac{1}{2}g_c\nabla^2 E - \frac{1}{2}\beta\nabla^2 E + \Theta(t - t_0)2C \cdot \cos(Kz - \Delta t)E \qquad (1),$$

where $E$ is the cavity field, $g$ is the gain, $I$ and $I_s$ are respectively the field and saturation intensities, $\alpha$ denotes the combined medium and waveguide losses, $g_c$ and $\beta$ are the gain curvature and dispersion, respectively. The fast-gain saturation is responsible for the stabilization of the system on a quasi-continuous intensity state throughout the whole cavity. The RF gain modulation translates to phase modulation through the linewidth enhancement factor[46,47]. We keep only the phase modulation with depth $2C$ and resonant spatial frequency $K$, where $\Delta$ is the detuning relative to the free spectral range of the cavity, and $\Theta(t)$ is the switch-on step function.

To describe the system through synthetic lattice dynamics, we use the fact that our system has well-defined spatial modes with spacing $\Omega$, i.e., we can write the cavity field in the modal basis, $m$ using $E = \sum A_m e^{-im\Delta t - imKz}$. By plugging this ansatz into linear part of Eq. (1), we obtain the Hamiltonian for $t > t_0$

$$H_{lin} = \sum_m (V(m) - iGm^2)a_m^\dagger a_m + C(a_{m-1}^\dagger a_m + a_{m+1}^\dagger a_m) + i(g - \alpha)a_m^\dagger a_m \qquad (2),$$

where the dispersion acts as an on-site potential energy $V(m) = Dm^2 + \Delta m$ and the gain curvature as onsite losses $-Gm^2$, where $D = \frac{1}{2}\beta K^2$ and $G = \frac{1}{2}g_c K^2$, and the operators $a_m^\dagger$ and $a_m$ denote the creation and annihilation of a photon in mode $A_m$. We will later show that the contribution of the missing nonlinear part, the fast-gain, to the dynamics leads to an effective equal population of reciprocal space of the synthetic lattice, i.e. the cavity space. We define the $m = 0$ mode as the initial lasing mode. At $t_0$, we turn on the modulation, and effectively induce nearest-neighbor coupling between the modes equal to $C$. Note that depending on the detuning $\Delta$ the on-site potential is tilted akin to the effect of an electrical field, see Fig. 1B [18,29,44]. The last terms are onsite gain, $g$, and loss, $\alpha$.

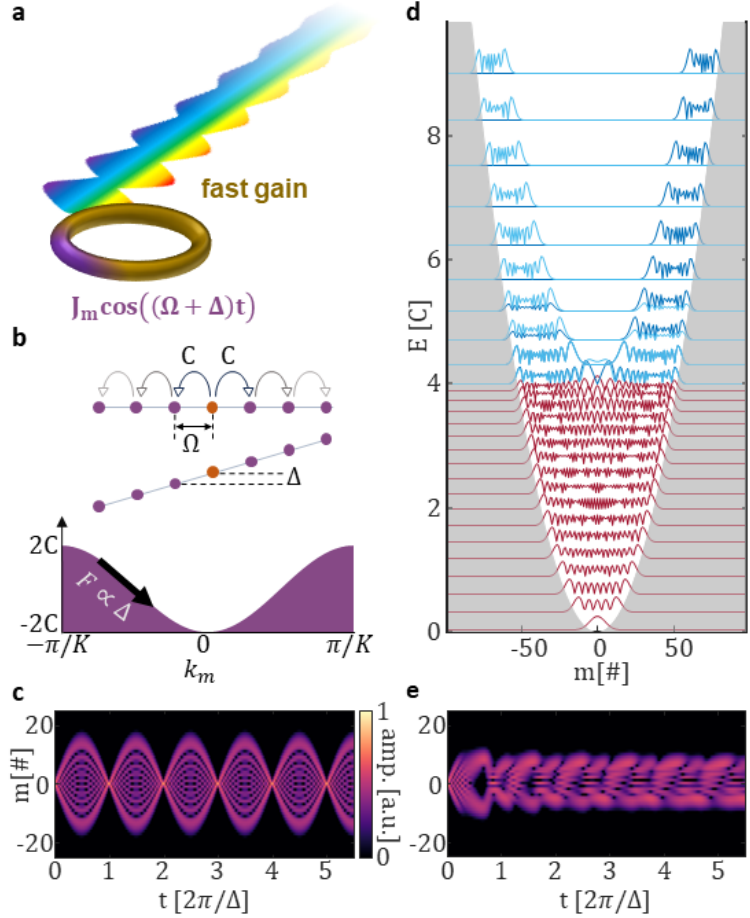

Figure 1. The fast-gain laser as a platform for the realization of nonlinear synthetic dimensions. (a) Fast-gain (gold) ring laser, with a section (purple) modulated at the resonance frequency of the cavity $\Omega$, with detuning $\Delta$ and depth $J_m$, emitting a multimode spectrum [multicolor beam, cf. Eq. (1)]. (b) Synthetic dimension lattices composed of coupled free spectral range modes of a perfect cavity. The modes are parametrically coupled by modulation, leading to effective hopping $C$, where detuned modulation leads to an on-site energy tilting (electric field) $\Delta$ [cf. Eq. (2)]. The resulting band structure inherits its shape from the modulation, with an effective semi-classical force proportional to the detuning. (c) Bloch oscillations that result from a single site excitation, when dispersion is absent. (d) The eigenmodes diagram in a quadratic potential induced by the dispersion, presented on a vertical energy scale; only every 4th mode is presented. Modes with $E > 4C$ become degenerate (marked with light and dark blue to differentiate the modes) but with opposite central momenta in the reciprocal space, taking the form of a Wannier-Stark ladder. (e) same as (c), but with complex dispersion; the oscillations are corrupted by the unequal energy spacing and selective dissipation.

**Artificial electric field and dispersed Wannier-Stark states**

Off-resonant modulation , i.e. $\Delta \neq 0$, is expected to produce an artificial electric field in the synthetic modal space[18,30]. In the absence of complex dispersion ($D = G = 0$), as the electric field increases, transport becomes less probable, and the extended Bloch states morph into a Wannier-Stark ladder, where the electron trajectories become more localized[48,49]. In a semiclassical analogue, we quantitatively analyze the Bloch periodicity through $k_m$, the reciprocal coordinate to the modal space $mK$, which experiences an effective linear force so that $\langle \dot{k}_m \rangle = F = -\Delta/K \rightarrow \langle k_m \rangle K = Kz - \Delta t$. In the absence of dispersion, the band structure is $E(k_m) = 2C \cos(k_m K)$ (Fig. 1B), so that the motion in the frequency ladder would have an effective velocity of $\langle \dot{m} \rangle K = -2CK \sin(\langle k_m \rangle K)$, with a period of $T_{osc} = \frac{2\pi}{\Delta}$ and oscillation magnitude $2C/\Delta$. Clearly, $k_m$ is directly related to space in the corotating frame, $z$, meaning that by choosing the RF modulation, we directly dictate the shape of the band structure. Initialized in a single site, the reciprocal space would be fully populated and therefore the wavefunction would be performing Bloch oscillations of the type presented in Fig. 1C. Even in the presence of the fast-gain saturation, the system would produce the same dynamics as in Fig. 1C, as the population in the reciprocal space is only shifting without changing its density pattern. This radically changes in the presence of a quadratic potential.

The first order dispersion sets a quadratic potential in the synthetic lattice, which is expected to exist in most observations of Bloch oscillations in doped semiconductors, as charges induce a Hartree potential in the biased lattice[50]. This breaks the translational symmetry of the problem, and calls for a general eigenmode analysis[51]. The eigenstates of the linear and Hermitian synthetic lattice ($G \to 0, g, \alpha \to 0$) echo the competition between the parabolic confinement, $Dm^2$, and kinetic energy, $4C$, see Fig. 1D. Note that the structure of the solution remains qualitatively the same for any $\Delta$. Within the kinetic energy bandwidth, $E_K = 4C$, the excitations explore the extent of the parabolic confinement to form Hermit-Gauss type solutions, see Fig. 1D. By equating the kinetic energy contribution of the coupling terms to the potential energy $Dm^2$, we can find the maximal site number of the Hermit-Gauss states to be $m_c = 2\sqrt{C/D}$. Above $E_K$, the excitations are localized around the sites where the on-site energies overtake the kinetic term; these states are similar to Wannier-Stark. It is possible to initialize the system in a single mode that overlaps with the Wannier-stark like ladder, with $m > m_c$. Fig. 1E shows a simulation of such dynamics in the absence of fast gain. Although the interference of perfectly equally spaced Wannier-stark modes should produce Bloch oscillations, the dispersion of our system corrupts the perfect spacing and ruins the periodicity of the motion. Moreover, the spectrally dependent losses play an additional destructive role for the observation of the oscillations. To study this mode competition in our system, we recall that the initial state is populating solely $m = 0$ with initial energy $E(0) = 0$ (Fig. 2A). The detuning, $\Delta$, translates the minimum of the potential by $m_\Delta = \frac{\Delta}{2D}$ and lowers it by $E_\Delta = \frac{\Delta^2}{4D}$ (Fig. 2B), effectively equipping the initial state at $m = 0$ with this relative energy. When this energy is smaller than the kinetic energy, the state falls into the Hermit-Gauss bound modes of the system (red region in Fig. 2A). Therefore, the minimal detuning required to "ionize" the state by exceeding the limit of $E_K$ is the critical value of $\Delta_c = 4\sqrt{CD}$, see Fig. 2B.

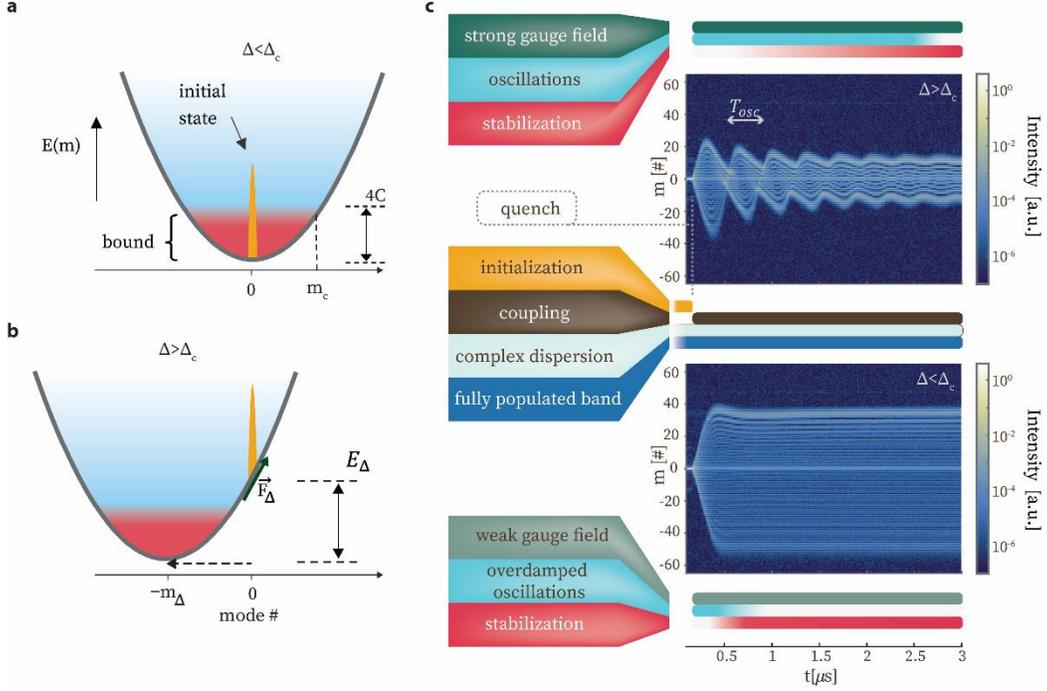

*Figure 2. Enabling oscillations in complex dispersion through the fast-gain emulator - a tool to study collective non-equilibrium dynamics. (a) The initial state is located at the central mode $m = 0$. When switching the parametric-coupling on with $\Delta = 0$ (top), we expect to observe only bound spectral states due to the parabolic confining potential (this is also the case when $\Delta$ is too weak). Once the detuning is large enough with $E_\Delta > E_K$ (b), the initial state excites modes that are mostly bound to the onsite quadratic energy, allowing the state oscillatory motion in the modal space. (c) Measurement of dispersive Bloch Oscillations and their stabilization. Left column presents functions and monitors of the emulator. On the righthand side, we show time-resolved dynamics in the synthetic lattice, for modulation detuning $\Delta$ above and below the critical value $\Delta_c$. In both cases, we initialize the system in a single synthetic lattice site and at time $t_0 = 150 ns$ quench the state to a coupled and biased lattice with complex dispersion. Throughout the experiments, the fast-gain nonlinearity is restricting the reciprocal space of the synthetic lattice to be full (quasi constant intensity), leading to coherent dynamics. When $\Delta f = -3.02$ MHz, we observe oscillations with periodicity near $T_{osc} = 2\pi/\Delta$, unlike the corrupted dynamics expected from systems with complex dispersion (see Fig. 1E). This value of detuning is above $\Delta_c$. In the case of modulated $\Delta f = -1.22$ MHz (top) which is above $\Delta_c$, the spectrum expands ballistically, performs a damped oscillation, and stabilizes on a broad coherent spectrum.*

**Active photonic quench dynamics emulators**

We study the process of "ionization" and the mode competition in our fast-gain ring laser platform by performing a time-resolved measurement of the spectrum, effectively observing the evolution of the density of light in the synthetic lattice (Fig. 2C). In these measurements, we initialize the system in a single mode state; at $t_0$, we start to modulate the semiconductor laser with a frequency detuned from the RF resonance by $\Delta$, effectively quenching the state. For sufficient detuning $\Delta > \Delta_c$, we observe that the state oscillates with evident periodicity and then coherently stabilizes (Fig. 2,top), unlike the expectation of oscillation decoherence in complex potentials (Fig. 1E). However, by detuning the modulation less than $\Delta_c$ we observe rapid expansion of the state, followed by a damped oscillation, and subsequent stabilization on a broad steady state (Fig. 2,bottom). This behavior corresponds to bound mode dynamics, where the added kinetic energy due to the dispersion shift is not sufficient to ionize the states. We compare measurements to our simulations and find excellent agreement (see section 7 in the SI). In doing so, we can retrieve a dispersion of ~$392 fs^2/mm$ and phase modulation depth per unit of time $M \approx 1.97 \cdot 10^8 \frac{rad}{s}$ to find the critical ionization frequency at $\Delta f_c = \frac{\Delta_c}{2\pi} \approx 2.64\ MHz$.

We ascribe the existence of oscillations in Fig. 2C to the fast-gain nonlinearity term, $g(1 - I(t,z)/I_s)E$, which is responsible for the stabilization of the system on a quasi-continuous intensity state in the cavity, restricting the number of photons, similar to hardcore blockade[52,53]. In fact, the timescale of the fluctuation suppression induced by a fast dissipative 4-wave mixing process[54,55] is faster than other dominant timescales in the system – coupling and complex dispersion. Thereby, at all times, the fast saturation mechanism is effectively restricting full population of the reciprocal space of the synthetic lattice (i.e. the cavity), overcoming the complex dispersion and reenabling oscillation dynamics (see sections 2 and 8 of the SI). This fast interaction, which translates into a long-range nonlinearity in the synthetic lattice, is giving the light in the system a liquid-like property[39], where the fast gain is dumping every density fluctuation in the reciprocal space (effectively the cavity space). This causes the reciprocal space to be equally occupied, similar to Fermi-Dirac statistics when the Fermi level permits.

**Characterizing the quench dynamics**

We analyze the oscillations and their stabilization rate after quenching to the dispersed Wannier-Stark ladder, with the shifted mode structure by $m_\Delta$. We compare several cases of spectral evolution with modulation detuning of $\Delta/2\pi = \Delta f = 4.22 MHz, 6.22 MHz, 8.22 MHz$, where increasing the modulation frequency reduces the period of the oscillations and its magnitude, as expected, see Fig. 3A. We observe that the oscillation period deviates from the perfect Bloch period $T_{osc}$; such change emerges mainly from the underlying quadratic potential . Figure 3B shows the calculated oscillation frequency of the expansion and contraction of the spectrum for different values of dispersion with ratios $r_D = D/D_{exp} = 0.7, 1.05, 1.4$, while $D_{exp}$ is the extracted dispersion from experiments. The periodicities extracted from the measurements in Fig. 3A are plotted as well. We find that relatively far from resonance, the oscillation frequency, $\Omega_{BO}$, is following the detuning $\Delta$, however, as the detuning approaches the critical value $\Delta_c$, $\Omega_{BO}$ deviates from $\Delta$, indicating the presence of dispersion and nonlinear fast-gain. The calculated decay extracted by the contraction of the broadest spectra in each Bloch oscillation is plotted in Fig. S4 in the SI, where we find that smaller detuning produces also a slower decay, whereas higher detuning reaches a value of ~$1\mu s$ for all dispersion values. This means that although the gain is fast, the interplay between the gain and the dispersion is relatively slow, operating on a long (~$\mu s$) time scale. The stabilization process can be attributed to the interplay between the gain curvature and the ultrafast gain saturation that suppresses any fluctuation of the laser intracavity intensity in space and time, therefore preferring a single Wannier-Stark state that is spatially extended throughout the whole cavity (see section 7 in the SI). This stabilization dynamics is shown in Fig. 3C, showing the initial mode overlap of $\psi(0^+)$ for $\Delta f = 8.22 MHz$, and the evolution of the projection of state $\psi(t)$ on the system modes, displaying decay to one Wannier-Stark like state. During the decay process, the modes show an interference pattern, suggesting a coherent interplay throughout.

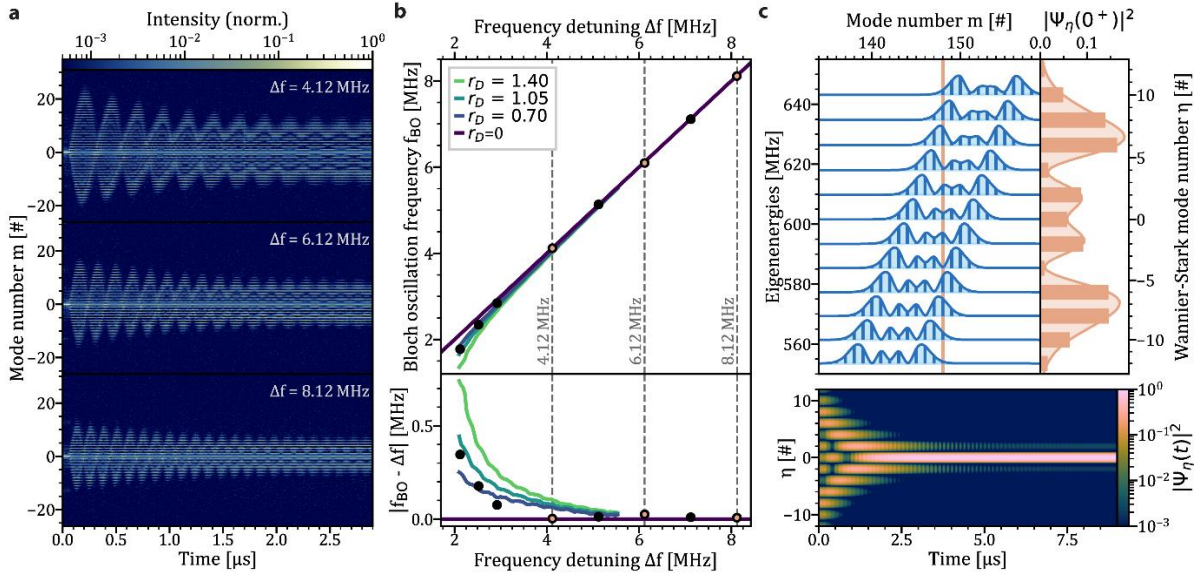

*Figure 3. Observation of Bloch oscillations in the synthetic space emulator and their decay to a Wannier-Stark state. (a) Spectral time resolved measurement at three different detuning values, $\Delta f = -4.22 MHz, -6.22 MHz, -8.22 MHz$. (b) Oscillation frequency $f_{BO}$ as a function of the detuning $\Delta f$ (top), and the oscillation frequency difference $\delta f = f_{BO} - \Delta f$ (bottom) for values extracted from (a) and different dispersion values as calculated from simulations. The oscillation frequency deviates from the modulation frequency for small detuning values due to increasing dispersion in the Wannier-Stark ladder that stems from the quadratic potential. We denote by $r_D$ the ratio between the dispersion value we used in each case and the experimentally retrieved value. (c) The initial state (orange line), projected on the Wannier-Stark supermodes, evolves in time to a single*

Attributing this decay rate to inherent characteristics of the gain in the system, we perform an additional time resolved measurement to investigate these contributions: we initialize the emulator in a state of very broad occupation of the synthetic lattice by resonant modulation, and then at time $t = 10.15 \mu s$, we turn off the modulation almost instantly (see section 6 of the SI), effectively quenching the state to the uncoupled lattice (Fig. 4A). Due to the absence of nearest-neighbor coupling through phase modulation, the fast-gain nonlinearity is not sufficient to maintain the broad state, and we observe a decay to $\delta(m)$ (Fig. 4A). Fig. 4B presents the decaying bandwidths of the measured spectrum over time, along with simulations at varying modulation depths. The decay times are in the range of $0.33 - 0.79 \mu s$, signifying the long time in our system related to the interplay between complex dispersion and the fast-gain nonlinear interactions. We notice that although the linear coupling between the modes is missing, the nonlinear long-range interaction induced by the fast-gain is maintaining coherent interference between modes, producing a Bessel-shaped state dynamics (Fig. 4C). This measurement demonstrates the uniqueness of the fast-gain, i.e., through the suppression of fluctuations and under substantial sub-microsecond changes it forces the system to maintain coherence between the modes throughout the whole evolution, representing the liquid phase in our emulator.

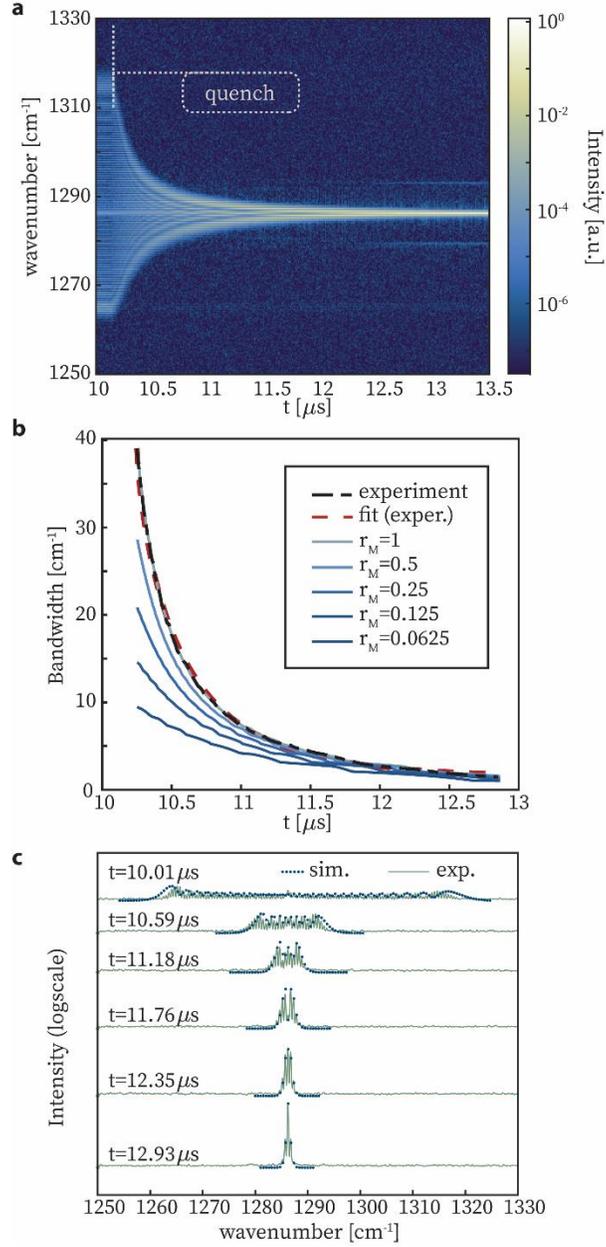

*Figure 4. Decay of a broad spectra when modulation is suddenly stopped. (a) We initialize the lasing system in a state with broad occupation in the synthetic lattice and quench it at $t = 10.15\mu s$ onto the uncoupled system. The initially broad lasing spectra, without a broadening mechanism such as the phase modulation, decays into a single mode state chosen by the gain curvature. (b) Calculated bandwidth vs. time for the measurement (with fit) and simulations of the decay rates for different relative values of the modulation before the quenching. (c) shows the Bessel shaped spectra throughout the decay, in the experiment (green line) and simulation (blue dots). The fast gain governs the dynamics by keeping coherence between the modes until full decay, preserving a Bessel-like state.*

**Concluding remarks**

In conclusion, we have demonstrated that the fast gain laser can support dynamics in a synthetic one-dimensional lattice with a parabolic potential, akin to the problem of a Hartree potential of a background doping. The quadratic potential limits Bloch oscillations up to sufficiently large electric fields applied to overcome the confinement of the potential well. Owing to fast-gain, the oscillations are seen in terms of the periodic population of the sites at a period that is close to the Bloch frequency, in spite of the complex potential imposed by the dispersion and gain curvature. Specifically, we found that, unlike a linear and translation-invariant system, the oscillation frequency deviates from the Bloch frequency as the detuning is reduced. Moreover, as the measurement proceeds, the oscillations decay on microsecond timescales to a single Wannier-Stark mode, a phenomenon rarely accessible with direct experiments in electronic systems[49]. We identify three timescales in our system, the longest is related to dispersion and linear dissipation, the intermediate is the Bloch oscillation period, and the shortest is fast-gain, which dictates a liquid phase for the light. The coherent transformation during quenching relates this process to an out-of-equilibrium liquid of light, having the small fluctuations decay faster than any other process. Intriguingly, the tendency of fast-gain systems to emit in quasi-constant intensity mirrors the influence of the Pauli exclusion principle on fermion population in lattices, thereby inhibiting condensation and the generation of optical pulses. Just as the fermionic nature of electrons compels them to populate the entire Brillouin zone, the fast gain is discouraging condensation of light intensity and evenly spans the state over the whole intracavity periodic[37]. Therefore, fast gain lasers are a versatile platform for solid state emulation of collective electronic phenomena, specifically, with interactions that lead to equalization of the density in the reciprocal space. As a platform, this system fulfills the requirements for a solid state emulator: system initialization, setting potential energies and supporting gauge fields, but with a unique mechanism that forces the probability density to be uniform in the band. Our platform is posited to enhance our understanding collective electron nonequilibrium dynamics within crystals, but also inspire new multifrequency photonic devices.